%% file: EOS-populations-paper.tex
\documentclass[aps,prd,twocolumn,nofootinbib,unsortedaddress]{revtex4-1}

\usepackage{aas_macros}
\usepackage{ns_bimodal}
\usepackage{analytic_scaling_RMS}

\usepackage{amsmath}
\usepackage{amsfonts}
\usepackage{amssymb}
\usepackage{mathtools}
\usepackage{mathrsfs}
\usepackage{bm}
\usepackage{graphicx}
\usepackage{xspace}

\newcommand{\param}{x}
\newcommand{\Param}{X}

\usepackage{xcolor}
\newcommand\editremark[1]{{\color{red} #1}}
\newcommand\SkipPreliminary[1]{}
\newcommand\SkipCrapAtEnd[1]{}

\definecolor{todocomment}{HTML}{D62728}
\definecolor{dwcomment}{HTML}{1F77B4}
\definecolor{rocomment}{HTML}{9467BD}
\definecolor{lwcomment}{HTML}{FF7F0E}
\definecolor{jrcomment}{HTML}{2CA02C}

\newcommand\E[1]{{\left\langle #1 \right \rangle}}
\newcommand\unit[1]{\mathrm{#1}}
\newcommand\rate{\mathcal{R}}

\newcommand\mchirp{{\mathcal{M}_{\text{c}}}}
\newcommand\mc{\mchirp}

\newcommand\cqg{CQG}

\usepackage{data_macros}

\renewcommand{\BibitemShut}[1]{}

\begin{document}

\preprint{}

\title{
  Inferring the neutron star equation of state simultaneously with the
  population of merging neutron stars
}

\author{Daniel Wysocki}
\email{dw2081@rit.edu}
\affiliation{Rochester Institute of Technology, Rochester, New York 14623, USA}

\author{Richard O'Shaughnessy}
\affiliation{Rochester Institute of Technology, Rochester, New York 14623, USA}

\author{Leslie Wade}
\affiliation{Kenyon College, Gambier, Ohio 43022, USA}

\author{Jacob Lange}
\affiliation{Rochester Institute of Technology, Rochester, New York 14623, USA}

\date{\today}

\begin{abstract}

   Observations of the properties of multiple coalescing neutron stars will simultaneously provide insight into neutron star mass and
  spin distribution, the neutron star merger rate, and  the nuclear equation of state.   Not all
  merging binaries containing neutron stars are expected to be identical.  Plausible sources of diversity in these
  coalescing binaries can arise from a broad or multi-peaked NS mass
  distribution;  the effect of different and extreme NS natal spins; the possibility of  NS-BH mergers; or even the possibility of phase transitions,  allowing for NS with
  similar mass but strongly divergent radius.   In this work, we provide a concrete algorithm to  combine all
  information obtained from GW measurements into a joint constraint on the NS merger rate, the distribution of NS properties, and  the nuclear equation of
  state.    
Using a concrete example, we show how biased mass distribution inferences can significantly impact the recovered
equation of state, even in the small-$N$ limit.  With the same concrete example, we show how small-$N$ observations
could identify a bimodal mass and spin distribution for merging NS simultaneously with the EOS.
Our concordance approach can be immediately generalized to incorporate other observational
  constraints. 

\end{abstract}

\maketitle

\section{Introduction}
\label{sec:intro}

The nuclear equation of state (EOS)---the relationship between pressure and density in cold nuclear matter---remains
weakly-constrained via terrestrial experiments, with differences having substantial impact on the
properties of  neutron stars  \cite{2016PhR...621..127L,2018RPPh...81e6902B}.   Conversely,  astrophysical observations of isolated and merging neutron stars provide a
natural mechanism to investigate the nuclear EOS.    For example, the size of isolated neutron stars is encoded in the pulsed or
bursty X-ray emission from their surface, allowing  observations and theoretical modeling of galactic
X-ray sources to
limit the range of possible neutron star mass-radius relationships \cite{2010PhRvD..82j1301O,2010ApJ...722...33S,2014ApJ...784..123L,2016RvMP...88b1001W,2019arXiv191211031R,2019ApJ...887L..24M}.
Neutron stars in coalescing binaries are subject to strong tidal interactions in the late stages of inspiral, which have
an observationally accessible impact on the outgoing gravitational wave signal
\cite{2008PhRvD..77b1502F,2014PhRvL.112j1101F} and thus enable constraints on the nuclear EOS \cite{2013PhRvL.111g1101D,2015PhRvD..92b3012A,2015PhRvD..91d3002L}.   With GW170817, the  imprint of these tidal interactions on the inspiral
signal was first constrained \cite{LIGO-GW170817-SourceProperties,2018PhRvL.121i1102D}, with widely-investigated follow-on implications for the nuclear
equation of state   \cite{LIGO-GW170817-EOS,2015PhRvD..91d3002L,2018PhRvL.120z1103M,2018PhRvC..98c5804M,2017ApJ...850L..34B,2017ApJ...850L..19M,2018arXiv181012917R,2018MNRAS.480.3871C,2018arXiv181204803C,gwastro-nucmatter-CapanoTewsEtAl2019,2019arXiv191009740E,LIGO-GW170817-EOSrank}.  
As more coalescing binary neutron stars are discovered in the immediate future, similar GW observations will even more
tightly constrain the nuclear equation of state both alone (e.g., \cite{2015PhRvD..91d3002L,2018PhRvL.120z1103M,2018PhRvC..98c5804M}) and
in conjunction with electromagnetic observations (e.g., \cite{2017ApJ...850L..34B,2017ApJ...850L..19M,2018arXiv181012917R,2018MNRAS.480.3871C,2018arXiv181204803C}).

GW measurements of coalescing NS and BH will also determine  the rate at which binaries with specific parameters merge.
GW observations by Advanced LIGO \cite{2015CQGra..32g4001T}  and Virgo \cite{gw-detectors-Virgo-original-preferred} have identified a binary
neutron star merger  \cite{LIGO-GW170817-bns,LIGO-GW170817-SourceProperties}.
As  envisioned originally in prototype investigations (e.g., \cite{2010CQGra..27k4007M,2012CQGra..29l4013S})  and as now made concrete with specific analysis procedures
\cite{2010CQGra..27k4007M,2013PhRvD..88h4061O,gwastro-EventPopsynPaper-2016,2017ApJ...846...82Z,2018MNRAS.477.4685B,2017PhRvD..96f4025M,2019MNRAS.484.4216R,gwastro-PopulationReconstruct-Parametric-Wysocki2018},
 the population distribution can be inferred phenomenologically,
 by combining observations
while accounting for parameter-dependent detector sensitivity. %
In principle,  the nuclear equation of state adds  only a handful of parameters to an already-large
phenomenological space used to characterize a compact binary population.
binaries,  
In this work, we demonstrate how to construct  simultaneous inference on the NS population and the nuclear
EOS, and the  potential of this approach to improve future GW measurements of the nuclear EOS.    Concretely, 
building on previous work \cite{2015PhRvD..91d3002L,gwastro-PENR-RIFT,2018arXiv181112529L}, we present and provide a
general-purpose code to perform this inference.   Our code combines the techniques from Wysocki et al
\cite{gwastro-PopulationReconstruct-Parametric-Wysocki2018}  (for population modeling) with  Carney et al
\cite{2018PhRvD..98f3004C}'s implementation of Lindlbom's EOS representation \cite{2010PhRvD..82j3011L}.    In order to perform this inference hierarchically, we estimate and re-use marginal likelihoods.  The organization of our inference strategy has  much wider applicability, both to more
generic EOS parameterizations and to other astrophysical inference scenarios involving parametric dimensional reduction
(e.g., inference for a subpopulation of binary black holes with exactly zero spin).

Our approach does not rely on any assumed or approximate similarities between different neutron stars to draw conclusions from the whole population
(cf. \cite{2012mgm..conf..743M,2019arXiv190205078F,2019arXiv190204557K,2019arXiv190410002R})
nor do we adopt a fiducial NS population distribution
(cf. \cite{2013PhRvL.111g1101D,2015PhRvD..91d3002L,2019arXiv190408907M,2019PhRvD.100j3009H}). %
If indeed all coalescing NS are identical and easily discriminated from binaries involving BHs---or even if the
differences are present but substantially smaller than our measurement error---then the sophisticated techniques
described in this work aren't necessary to interpret the first few coalescing BNS.  As more binary NS accumulate,
however, the methods described in this work will be increasingly necessary to fully exploit all available information
and to enable high-precision measurements of correlated BNS properties. 
Too, the tidal parameters which influence the GW phase are not necessarily common to all events. 
  Plausible sources of diversity in these
  coalescing binaries can arise from a broad or multi-peaked NS mass
  distribution;  the effect of different and extreme NS natal spins; the possibility of  NS-BH mergers; or even the possibility of phase transitions,  allowing for NS with
  similar mass but strongly divergent radius.  
To the extent all coalescing NS are not identical, this kind of  approach will be required to infer the nuclear equation of
state even in the immediate future.

This paper is organized as follows.
In Section \ref{sec:method}, we review our framework for population inference in general and the nuclear
equation of state in particular.  We address  challenges
for efficient computation appropriate to models (like the nuclear equation of state) in which the population model predicts all objects occupy a
lower-dimensional subspace of the entire physical observable space.
In Section \ref{sec:results}, we demonstrate our method, recovering the nuclear equation of state from 
neutron stars generated from a bimodal mass and spin distribution, consistent with current observations.
Using a concrete counterexample, we show that inference of the mass, spin, and EOS must be performed simultaneously to
avoid introducing bias into the inferred EOS.
In Section \ref{sec:discussion}, we discuss our proof-of-principle calculation relates to our expectations about future
measurements.

\section{Method}
\label{sec:method}

\subsection{Population inference}
\label{sec:method:pop-inf}

In this section, we review the framework introduced in BPM \cite{gwastro-PopulationReconstruct-Parametric-Wysocki2018}
for population inference in general and the \textsc{PopModels} population inference code specifically,
modifying the notation to avoid collisions with the tidal deformability.
In the original BPM investigation, binaries coalesce at a spacetime-independent rate per unit comoving volume ${\cal R}$.  Binaries with
intrinsic parameters $x$ would merge at a rate $dN/dVdt dx = {\cal R} p(x)$.  
 The intrinsic parameters
$x$ that describe a binary in quasicircular orbit are the individual component  masses $m_i$ and spins
$\bm{S}_i$ ($i=1,2$) at some reference time.  We characterize compact object spins using the dimensionless variable 
$  \bm{\chi}_{i} = \bm{S}_{i}/m_{i}^{2}$.    We characterize the matter-dependent factors which influence point-particle
 motion by the  dimensionless tidal deformabilities $\Lambda_i$  \cite{2008PhRvD..77b1502F,2010PhRvD..81l3016H} (i.e., $\Lambda_i/m_i^5$ is the ratio between the NS
 induced quadrupole and the applied quadrupolar field).  We assume other  degrees of freedom like the quadrupole
 moment which enter into the orbital evolution are  well-determined in an EOS-independent manner by $\Lambda_i$, extending the Darwin-Radau
 and related approximations to neutron stars; see, e.g.,  \cite{2018PhRvD..97h4038P,2019arXiv190303909C} and references therein.
BPM  requires an estimate  $\mu(\Param)$ of how many events a given experiment should find on average. 
We follow previous work by estimating this rate $\mu$ using a characteristic sensitive volume, denoted $VT$.
For binary neutron stars, the sensitive volume depends principally on the binary chirp mass $\mc=(m_1 m_2)^{3/5}/(m_1+m_2)^{1/5}$, which for binary neutron
stars spans a small range.
In terms of these ingredients, BPM expresses the likelihood of an astrophysical BBH population with parameters $X$
as a conventional inhomogeneous Poisson process:
\begin{equation}
  \mathcal{L}(\rate, \Param) \propto
  e^{-\mu(\rate, \Param)}
  \prod_{n=1}^N
    \int \mathrm{d}\param \, \ell_n(\param) \, \rate \, p(\param\mid\Param),
  \label{eq:inhomog-poisson-likelihood}
\end{equation}
where $\mu(\rate,\Param)$ is the expected number of detections under a given population parameterization $\Param$ with
overall rate $\rate$ and where $\ell_n(x)=p(d_n|x)$ is the likelihood of data $d_n$---corresponding to the $n$th detection---given binary parameters
$x$.
The population inference code \textsc{PopModels} \cite{gwastro-PopulationReconstruct-Parametric-Wysocki2018}, which we employ and extend in this work, provides a set of building blocks with which to assemble very general $p(\param|\Param)$.  In the context of
this work,  we'll be interested specifically in Gaussian distributions (for mass); $\beta$ distributions (for spin magnitude); and mixture models for multiple sub-populations.  We will not
allow for  neutron star spin-orbit misalignment, being a highly subdominant effect for the NS spin magnitudes we expect.

In principle, Eq.~\ref{eq:inhomog-poisson-likelihood} can be used in any general-purpose Bayesian inference engine (e.g., direct
quadrature; MCMC) to perform simultaneous
inference on all  $d\times N+1+D$ relevant parameters, where $N$ is the number of observations, $d$ is the
individual-event model dimension size (here, approximated by $6$), and $D$ is the number of hyperparameters needed to
characterize the NS population (e.g., mass distribution, spin distribution, and equation of state).
In many fields, including previous efforts to infer the nuclear equation of state from X-ray binaries, this direct
approach is used; see, e.g., \cite{2010PhRvD..82j1301O,2010ApJ...722...33S,2014ApJ...784..123L}.
But the calculation can conceivably be reorganized  to efficiently and hierarchically re-use fiducial analyses of each
event, allowing much more rapid analysis and extension of results, essential given the computational cost of each event
in isolation and the number of events requiring analysis in the  immediate future.

One conventional approach for efficient hierarchical calculation  (see, e.g.,  BPM~\cite{gwastro-PopulationReconstruct-Parametric-Wysocki2018} and references
therein) performs a conventional Markov-chain Monte Carlo (MCMC) analysis for each event for all intrinsic and extrinsic parameters.
This fiducial analysis of each GW event, derived using a set of reference prior assumptions, requires an analysis with  all parameters $y=(x,\theta)$ needed to characterize the
quasicircular binary.  We  use the (Gaussian) likelihood function $p(d|y)$ for detector network data $d$ containing a
signal, and apply Bayes' theorem and some fiducial assumptions $p_{\rm ref}(y)$ to deduce the posterior distribution
$p(y|d)\propto p(d|x)p(y)$.  
Standard Bayesian tools~\cite{PEPaper,gw-astro-PE-lalinference-v1} will produce a sequence of independent, identically
distributed samples $x_{n,s}$ ($s=1,2,\ldots,S$) from the posterior distribution $p(x|d)$ for each event $n$.  
The integrals $\int dx \ell_n(x) p(\param|\Param)$ can then be performed via Monte Carlo, using the fiducial samples
provided by our reference analysis.
For this conventional approach to work, however, the model predictions $p(x|\Param)$ must not be a set of measure zero, like a
submanifold (e.g., binaries with exactly a specific value of spin, or binaries which have a deterministic mass-tide relation
$\Lambda=\Lambda(m)$).

Unfortunately,  EOS inference and other astrophysically-motivated questions involve dimensional reduction: their
formation model predicts a deterministic relationship between binary parameters.   For EOS inference and to the level of
accuracy discussed in this work, that deterministic relationship is  $\Lambda(m)$.  For astrophysical formation channels
which predict nearly-maximal or exactly-zero spins for binary black holes which undergo certain formation channels, that
relationship is a fixed value of the spin magnitude (for some black hole masses, some of the time).  Other formation
channels may predict infinitesimal BH spin-orbit misalignments for certain binary BH masses.    For all of these
questions, the straightforward hierarchical technique described fails.  If the space of low-dimensional scenarios is
finite, like a finite list of possible EOS or a finite set of BH natal spin scenarios, then inference could be carried
out for every combination of possibilities.   
But usually the model space is either infinite or large, and the overhead of carrying out repeated inference is
prohibitive.

\subsection{Individual event inference via marginal likelihood models}
\label{sec:method:likelihood}

To circumvent the problems with dimensional reduction identified above, building on previous work
\cite{2015PhRvD..91d3002L,gwastro-PENR-RIFT,2018arXiv181112529L}, we instead perform the integrals appearing in our
expression with  the (marginal) likelihood $\ell_n(x)$.     Some parameter inference engines like RIFT
\cite{gwastro-PENR-RIFT} already produce and export an estimate of the (marginal)
likelihood as a data product, using either Gaussian process or random forest interpolation. [For high-amplitude signals, the RIFT marginal likelihood can
  often be approximated by a Gaussian in suitable coordinates; see, e.g., \cite{NRPaper}.]  For conventional MCMC engines, which only report posterior samples, the likelihood can
sometimes be approximated by a well-tuned density approximation like a Gaussian kernel density estimate; see, e.g.,
\cite{2015PhRvD..91d3002L,2018arXiv181112529L,2019arXiv190408907M}.   Finally, for simple investigations which don't require end-to-end parameter inference,   a suitable
approximate marginal likelihood $\ell_n(x)$ can easily be generated using  a Fisher matrix approximation, as is
implemented in the \textsc{synthetic-PE-posteriors}
library \footnote{\texttt{https://git.ligo.org/daniel.wysocki/synthetic-PE-posteriors}} \cite{gwastro-PopulationReconstruct-Parametric-Wysocki2018}.

To complicate matters, for binary neutron stars, the marginal likelihoods $\ell_n(x)$ are exceedingly narrow relative to fiducial astrophysical
priors $p(x)$ and any plausible model predictions $p(x|X)$.  For example, observations of GW170817 constrained its (redshifted) chirp mass $\mc (1+z)$ to within $10^{-4} M_\odot$.  Hence the marginal integrals $\int dx
\ell_n(x) p(x|X)$  require event-specific adaptive sampling in $x$.   
We  modify the limits of each
integral  over chirp mass to conservatively contain the support of $\ell_n(x)$.

\subsection{EOS spectral decomposition}
\label{sec:method:eos-spectral}

In this work we adopt the spectral EOS parameterization introduced by Lindblom  \cite{2010PhRvD..82j3011L}, implemented
in Carney et al \cite{2018PhRvD..98f3004C} in \textsc{LALSuite} \cite{lalsuite},
and previously used to interpret GW170817 \cite{LIGO-GW170817-EOS}.  
In this specific representation, the nuclear equation of state relating energy density $\epsilon$ and pressure $p$ is
characterized by a low-density SLy EOS joined to a a spectral representation  
 at $p_0=5.4\times 10^{32}\unit{dyne}\; \unit{cm}^{-2}$, using a high-density 
four-parameter spectral model characterized by its adiabatic index $\Gamma(p)$:
\begin{eqnarray}
\ln \Gamma(p) = \sum_{k=0}^3 \gamma_k [\ln (p/p_0)]^k,
\end{eqnarray}
where $\gamma_k$ are expansion coefficients. %
  From the adiabatic index, the
equation of state follows via solving
\begin{eqnarray}
\frac{d\epsilon}{dp} = \frac{\epsilon+p}{p\Gamma(p)}.
\end{eqnarray}
From the pressure and energy density, other state variables can be calculated,
such as the baryon rest mass density $\rho_b=m_b n = (\epsilon+p)/e^h$, which follows from the
pseudo enthalpy $h$  via  $dh/dp=1/(\epsilon+p)$; see, e.g., the discussion in \cite{2009PhRvD..79l4032R}.
As a fiducial EOS, we will adopt the spectral approximation to APR4 from \textcite{2010PhRvD..82j3011L},
 given by $\gamma_0 = 0.8651$, $\gamma_1 = 0.1548$, $\gamma_2 = -0.0151$, $\gamma_3 = -0.0002$.

Because of the exponential dependence of $\Gamma$, only a narrow region for $\{\gamma_k\}$ produces observationally plausible
equations of state, and we place limits on $\{\gamma_k\}$ that are largely consistent with prior work \cite{2018PhRvD..98f3004C}.
 To be consistent with the wide range of proposed models, we require that $\Gamma\in[0.6,4.5]$.  
We require for simplicity that the EOS produces maximum NS masses greater than $1.97 M_\odot$; see, e.g.,
\cite{2019arXiv190408907M} for a more careful treatment of uncertainties in the observed maximum mass.   To allow for the
 possibility of causal EOS being approximated within our model family by an acausal representation, we
 require the inferred EOS is approximately causal (i.e., $v_s =\sqrt{dp/d\epsilon}< 1.1 c$) up to the central pressure of the most massive NS permitted by the EOS.
As in previous work, we assume the prior on $\gamma_k$ is a constant value as a function of $\gamma_k$,
 and adopt the prior ranges used in
previous work.
Since the region of equations of state allowed by the aforementioned criteria occupies a subset of the prior range on $\{\gamma_k\}$ which is not closely aligned with the coordinates $\{\gamma_k\}$, we initialize our MCMC with a rotated coordinate system, as described in Appendix \ref{ap:pca}.  The MCMC we employ is affine-invariant, so the rotated coordinate system will provide no sampling improvements post-initialization, but for samplers without affine invariance this rotated system will be very useful.%

\subsection{Source population model}

Motivated by observations of galactic binary neutron stars \cite{2016ARA&A..54..401O,2018MNRAS.478.1377A}, we explore a two-component population of neutron stars, with overall minimum and maximum masses set by the nuclear equation of state.  To emphasize the importance of an accurate model for the mass distribution we also employ a one-component population in our inferences, which cannot capture the full complexity of the two-component model we synthesized data from.  Specifically, we employ a mixture model for binary components $x = (m_1,m_2,\chi_{1,z},\chi_{2,z},\Lambda_1,\Lambda_2)$, defined as
\begin{gather}
  p(x) \propto
  \sum_{k=1}^K w_k \,
    \mathscr{M}_k(m_1) \, \mathscr{M}_k(m_2) \,
    \mathscr{S}_k(\chi_{1,z}) \, \mathscr{S}_k(\chi_{2,z})
  \label{eq:pop-model}
\end{gather}
everywhere that $m_{\mathrm{min}}(\boldsymbol{\gamma}) \leq m_2 \leq m_2 \leq m_{\mathrm{max}}(\boldsymbol{\gamma})$ and $\Lambda_i = \Lambda(m_i, \boldsymbol{\gamma})$, and zero elsewhere.  $\mathscr{M}_k$ and $\mathscr{S}_k$ represent the mass and spin distributions for the $k$th sub-population, respectively.  We model the $\mathscr{M}_k$'s as Gaussians with unknown mean and variance.  The $\mathscr{S}_k$'s are assumed to follow beta distributions bounded by $|\chi_z| < 0.05$, again with unknown mean and variance.  For simplicity's sake, we assert that the means and variances don't change between the primary and secondary NS.  The $\Lambda_i = \Lambda(m_i, \boldsymbol{\gamma})$ constraint introduces delta functions into the expression for $p(x)$, making it impossible to evaluate $p(x)$ numerically.  However, it is simple to draw samples from $p(x)$---we simply draw samples for ($m_1,\chi_{1,z},m_2,\chi_{2,z}$) and compute $\Lambda_i = \Lambda(m_i, \boldsymbol{\gamma})$ to produce corresponding samples for $\Lambda_1$ and $\Lambda_2$.  Informed only by the limited dimensionality of $\Lambda_i$, one could compute the integrals $\int \mathrm{d}x \ell_n(x) p(x|X)$ by drawing samples $x_{n,s}$ from $p(x)$ in this manner and computing $\frac{1}{N} \sum_s \ell_n(x_{n,s})$.  However, the very tight constraints on $\mchirp (1+z)$ make this require very high $N$ for the integrals to converge.  Instead, we separate the population's distribution into $p(x) = p(m_1, m_2) p(\chi_{1,z}, \chi_{2,z}, \Lambda_1, \Lambda_2 | m_1, m_2)$, and draw mass samples from a distribution $\mathbb{A}$ adapted to the region with non-vanishing $\ell_n(x)$, uniform in $\mchirp (1+z)$ and $\delta = (m_1 - m_2) / (m_1 + m_2)$.  The integral then becomes
\begin{equation}
  \int \mathrm{d}x \ell_n(x) p(x|X) \approx
  \frac{1}{N}
  \sum_{s=1}^S
    \frac{p(m_{1,s}, m_{2,s})}{\mathbb{A}(m_{1,s}, m_{2,s})} \,
    \ell_n(x_s),
\end{equation}
where
\begin{equation}
  \mathbb{A}(m_1, m_2) \,\propto\,
  \boldsymbol{J}^{-1}_{(m_1,m_2)\to(\mchirp,\delta)}(m_1, m_2),
\end{equation}
and $\boldsymbol{J}^{-1}$ is the inverse Jacobian determinant
\begin{widetext}
\begin{equation}
  \boldsymbol{J}^{-1}_{(m_1,m_2)\to(\mchirp,\delta)} =
  \begin{vmatrix}
    \frac{m_2 (2 m_1 + 3 m_2)}{5 (m_1 m_2)^{-2/5} (m_1 + m_2)^{-6/5}} &
    \frac{m_1 (2 m_2 + 3 m_1)}{5 (m_1 m_2)^{-2/5} (m_1 + m_2)^{-6/5}}
    \\
    \frac{2 m_2}{(m_1 + m_2)^2} &
    \frac{-2 m_1}{(m_1 + m_2)^2}
  \end{vmatrix}
\end{equation}
\end{widetext}

For our fiducial model, we took a two-component ($K=2$ in Eq.~\label{eq:pop-model}) mass distribution based on the
galactic neutron star constraints from \textcite{2018MNRAS.478.1377A}, in the first row of their Table 3.  Rather than
take their maximum \emph{a posteriori} values, we approximated their reported estimates as Gaussians, and took a single
draw, resulting in $\mathbb{E}[m]_1 = 1.34 \, M_\odot$, $\mathrm{Std}[m]_1 = 0.05 \, M_\odot$, $\mathbb{E}[m]_2 = 1.88
\, M_\odot$, $\mathrm{Std}[m]_2 = 0.32 \, M_\odot$, and relative weights of $6:4$.  For the low mass component's spin
distribution, we utilize a zero-spin model, attainable using a $\beta$ distribution with $\mathbb{E}[\chi_z]_1 = 0$ and
$\textrm{Std}[\chi_z]_1 \to 0$.  For the high mass component, however, we expect higher spins, as these would likely be
recycled pulsars \cite{2008LRR....11....8L}, and so we adopt fiducial choices $\mathbb{E}[\chi_z]_2 = 0.02$ and $\mathrm{Std}[\chi_z]_2 = 0.01$.

All of our analyses use uninformative uniform priors on the spectral EOS parameters, following previous work
\cite{2018PhRvD..98f3004C,LIGO-GW170817-EOS}; see Table \ref{tab:prior-EOS}, and  Appendix \ref{ap:pca} for more discussion.  Our fiducial analyses use uninformative priors, uniform in $\mathbb{E}[m]$, $\mathbb{E}[\chi_z]$, $\mathrm{Std}[m]$, and log-uniform in each sub-population's rate and $\mathrm{Std}[\chi_z]$; see Table \ref{tab:prior-fiducial}
To account for observations of galactic neutron stars, rather than reanalyze all galactic observations ourselves, we
employ pre-digested prior constraints on this same two-component mass model provided by  Table 3 of
\cite{2018MNRAS.478.1377A}.
In particular, we use an (improper) prior in the maximum neutron star mass $m_{\rm max}(\boldsymbol{\gamma})$,
extending from $1.97 M_\odot$ to infinity, to account
for the impact of the most recent well-determined NS masses on the inferred NS maximum mass  \cite{2013Sci...340..448A,2016arXiv160501665A,2019arXiv190406759C}.

\begin{table}
\centering
\begin{tabular}{p{0.25\columnwidth}p{0.25\columnwidth}p{0.25\columnwidth}p{0.25\columnwidth}}
  \hline
  $\gamma_0$ & $\gamma_1$ & $\gamma_2$ & $\gamma_3$
  \\ \hline
  $[0.2, 2]$ & $[-1.6, +1.7]$ & $[-0.6, +0.6]$ & $[-0.02, +0.02]$
  \\ \hline
\end{tabular}

\caption{
  Prior ranges for the spectral EOS parameters used in all of our
  analyses.  All priors are uniformly distributed.
}

\label{tab:prior-EOS}

\end{table}

\begin{table*}
\centering
\begin{tabular}{p{0.4\columnwidth}p{0.4\columnwidth}p{0.4\columnwidth}p{0.4\columnwidth}p{0.4\columnwidth}}
  \hline
  $\mathcal{R}$ [$\mathrm{Gpc}^{-3} \, \mathrm{yr}^{-1}$] &
  $\mathbb{E}[m]$ [$M_\odot$] & $\mathrm{Std}[m]$ [$M_\odot$] &
  $\mathbb{E}[\chi]$ & $\mathrm{Std}[\chi]$
  \\ \hline
  LU $[1, 10^5]$ &
  U $[0.9, 2.9]$ & LU $[0.05, 5]$ &
  U $[-0.05, +0.05]$ & U $[0, 0.05]$
  \\ \hline
\end{tabular}

\caption{
  Prior types and ranges for our fiducial analyses.  U denotes a
  uniform prior, and LU denotes a log-uniform prior.
}

\label{tab:prior-fiducial}

\end{table*}

\section{Results}
\label{sec:results}

\begin{figure}
\includegraphics[width=\columnwidth]{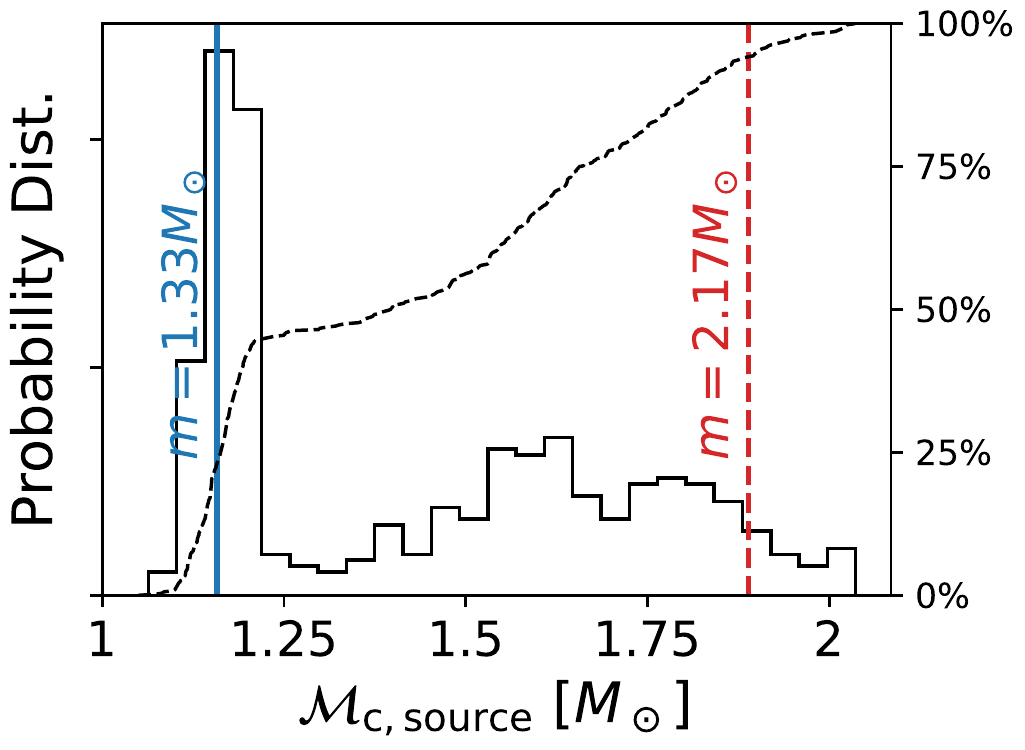}
\caption{
Synthetic detected observations:
Histogram (solid) and cumulative distribution (dashed) of chirp masses $\mc$ for
the synthetic observed population.  For reference, the leftmost vertical line
shows the chirp mass corresponding to two NS each of mass $1.33 M_\odot$.  The
rightmost vertical line shows the corresponding limit for two $2.17 M_\odot$ NS,
close to both the largest observed value and the maximum mass limit for
commonly-discussed equations of state.
}
\label{fig:synthetic}
\end{figure}

To illustrate our method, we generate a synthetic population of neutron stars drawn from our
fiducial bimodal population.  Assuming merging neutron stars are uniformly distributed in comoving volume and using a
naive detection model -- a single-interferometer SNR threshold of $8$ --  based on advanced LIGO's target sensitivity
(\textsc{aLIGODesignSensitivityP1200087} from \cite{LIGO-2013-WhitePaper-CoordinatedEMObserving};see \cite{LIGO-aLIGODesign-Sensitivity-Updated}), we construct a
population of {\nsynthetic} synthetic observations.  
As illustrated by Figure \ref{fig:synthetic}, the true parameters of this detection-weighted sample
include a fraction ($\simeq 10\%$) of events close to our presumed maximum NS mass. Our population inference thus constrains the
nuclear equation of state both through the maximum observed mass and through direct measurements of NS
tidal deformability.
Using RIFT on each observation $n=1 \cdots {\nsynthetic}$, we perform Bayesian inference to construct the
marginal likelihood $\ell_n(x)$ as a function of $x=(m_1,m_2,\chi_{1,z},\chi_{2z},\Lambda_{1},\Lambda_2)$, assuming each
source has an otherwise-determined sky location and redshift.
We performed parameter inference rapidly and in large scale, requiring subsequent hand-removal of some events suffering from convergence issues.
We construct several randomly-selected subsets of these {\nsynthetic} synthetic events, to generate synthetic observing scenarios
for the first 1, 5, and 10 coincident observations. 
Using  \textsc{PopModels} on each set of observations, we infer the EOS ($\boldsymbol{\gamma}$) and population hyperparameters ($\Param$), adopting a network with
presumed HLV design sensitivity.  In this work, we scale the  fiducial  analysis interval
of $7.22$ days to the number of events in our synthetic sample.
Note that due to selection effects, our synthetic population produces  roughly equal numbers of
observations from both components; see Figure \ref{fig:synthetic}.

\begin{figure*}
\includegraphics[width=0.4\textwidth]{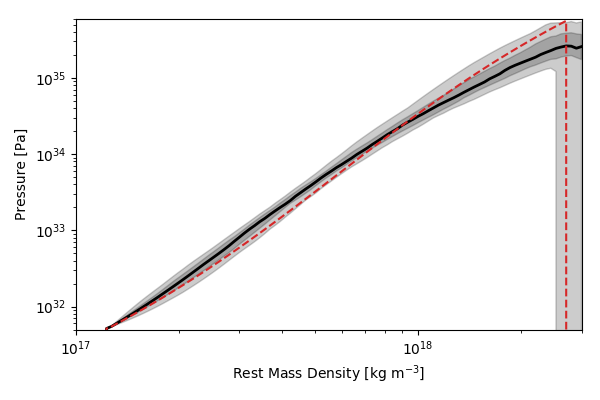}
\includegraphics[width=0.4\textwidth]{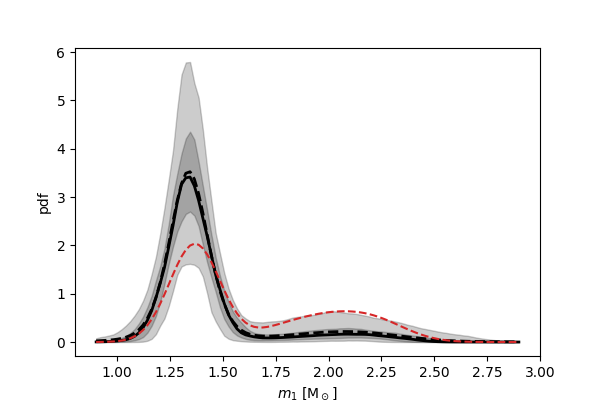}
\caption{
  Example of inference about NS mass, spin, and EOS, using 10 detections:
  \emph{Left panel}:
  Median (solid line) and 50\% and 90\% credible regions (shaded regions) for
  the nuclear equation of state.  True injected EOS overlain (red dashed line).
  \emph{Right panel}:
  Same as left, but showing the recovered neutron star mass distribution.
  Posterior predictive distribution also included, as dashed black line.
}
\label{fig:results}
\end{figure*}

\begin{figure*}
\includegraphics[width=\textwidth]{%
  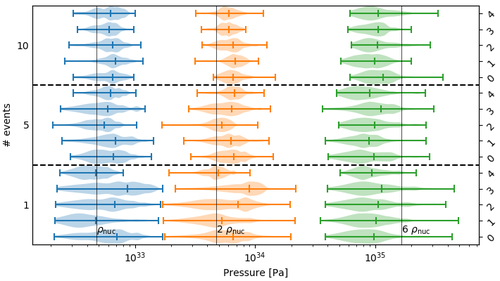%
}
\includegraphics[width=\textwidth]{%
  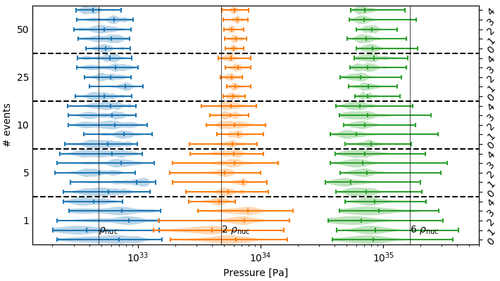%
}
\caption{
Posterior inferences about the pressure at three fixed densities.  The top panel
shows results from inference using a two-component mass model, consistent with
the underlying population.  At low density, GW observations only slowly improve
our understanding of the EOS.  At twice nuclear density, direct GW constraints
on tides inform the nuclear EOS.  At high density, GW observations provide less
new information.  The bottom panel shows how EOS biases arise by adopting a
unimodal population model which fits the NS mass distribution poorly,
emphasizing the need for careful mass distribution modeling.
\label{fig:results:eos_central}
}
\end{figure*}

Figure \ref{fig:results} shows inferences deduced from one of our synthetic 10-event populations.  We recover the injected EOS, identify both
populations in the NS mass distribution, and place weak constraints on NS spins.  
As the number of events increases, all our observational constraints become tighter, albeit strongly dependent on how well measured the added events are, and the particular properties of those events.
For such a simple Gaussian model, as discussed below quantitatively, the systematic and statistical accuracy to which we recover the
mass distribution can be understood by simple frequentist arguments (e.g., the accuracy in the measured mean mass of
each component). 
Less obvious and much more variable are our inferences about the EOS.
Figure  \ref{fig:results:eos_central} shows our results for the NS EOS at three fiducial densities.
The tidal deformabilty of NS correlates  with the central
densities of observed NS.   However, barring unlikely signal amplitudes, GW measurements of tides will constrain these deformabilities only when $\Lambda$ is
relatively large and thus the NS mass is relatively small.      Conversely, confident identification of NS with  large chirp masses
will require the high-density EOS produce NS with correspondingly large masses.   In our synthetic population, however,
such NS binaries occur rarely, with less than than 10\% of mergers providing meaningful new constraints on the maximum
NS mass.
For these reasons, observations of our synthetic population must most tightly   constrain the pressure at $\sim 2\rho_{\rm nuc}$.

\subsection{Understanding EOS constraints}

\begin{figure*}
\includegraphics[width=0.35\textwidth]{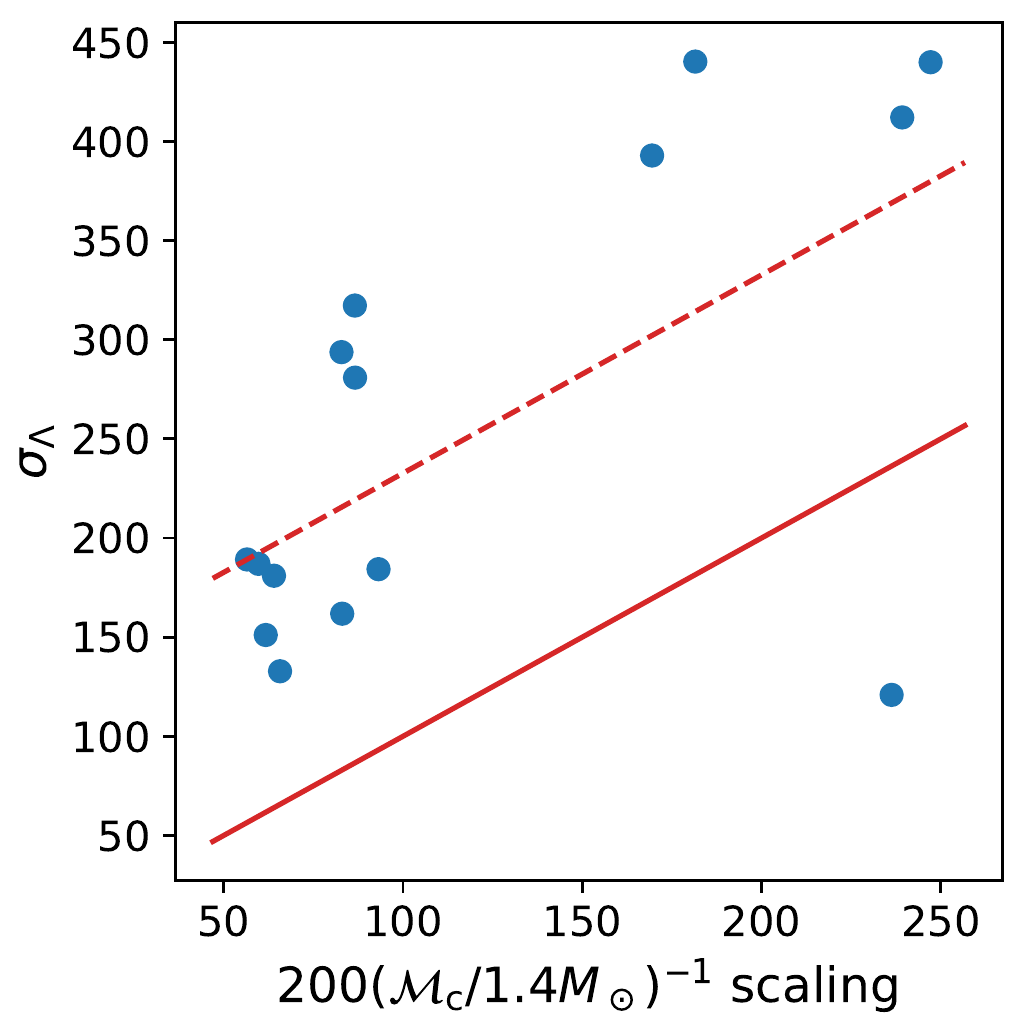}
~
\includegraphics[width=0.6\textwidth]{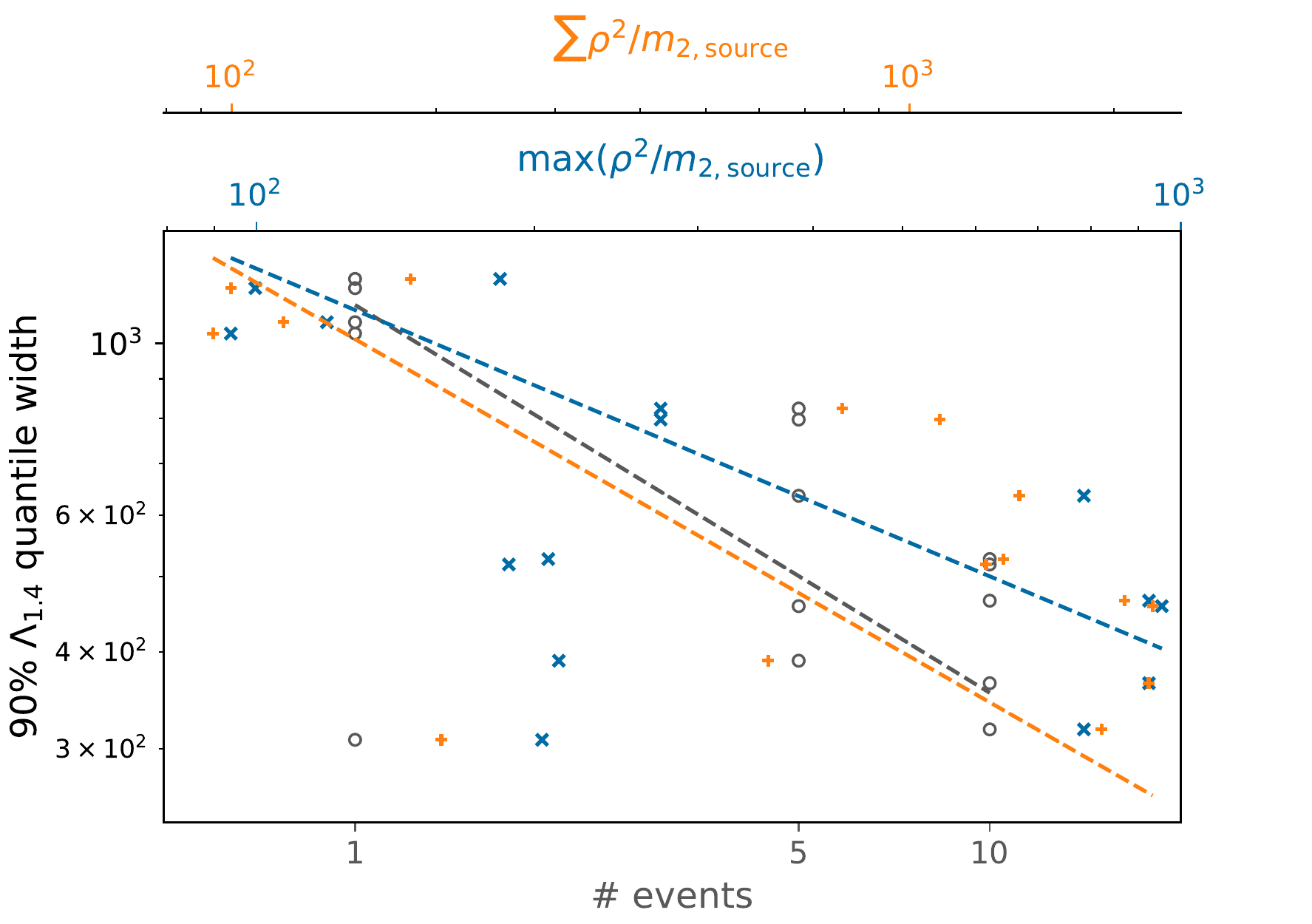}
\caption{
  EOS precision scaling:
  \emph{Left}:
  Comparison between the standard deviation on the derived parameter
  $\Lambda_{1.4}$ and predictions from the scaling relation described in
  {Sec.}~\ref{sec:results}.  If scaling were accurate, points would lie along
  the solid red line.  Dashed line represents same scaling, but shifted to
  intersect the first point along the $x$-axis.
  \emph{Right}:
  The points here show how our 90\% measurement error on $\Lambda_{1.4}$ scales
  against the number of events, as well as the most influential detection, and the
  sum of the influence of multiple detections, as is described in
  \ref{sec:discussion}.
}
\label{fig:results:HyperparametersVsIteration}
\end{figure*}

Our stacking strategy for multiple populations can be usefully compared with a more frequently discussed and much simpler scenario: where all
binary neutron stars have similar masses and hence tidal deformabilities. 
Previous studies have shown that the measurement error $\sigma_{\Lambda} \simeq 200(\mc/1.4 M_\odot)^{-1}$
 depends weakly on mass
 (see, e.g., Figure 7 of \cite{gwastro-nucmatter-CarsonSteinerYagi2019}).  Taking this scaling relation for
 $\sigma_{\Lambda}$, and adding its inverse in quadrature for multiple events ($\sigma_{\Lambda,\mathrm{tot}}^{-2} =
 \sum_k \sigma_{\Lambda,k}^{-2}$), we find that this scaling relation roughly holds, but that it must be shifted to
 higher errors, as is shown in {Fig.}~\ref{fig:results:HyperparametersVsIteration}.  We find an RMS error for this
 shifted relation at {\RMSforMC}.

Two of the largest driving factors for a BNS's contribution to measuring the EOS are its signal-to-noise ratio, $\rho$,
and the mass of the smallest object, $m_{2,\mathrm{source}}$.  So while our stacking method should reduce the
uncertainty on $\Lambda_{1.4}$ with each detection, not all detections are created equally.  As illustrated in Figure
\ref{fig:results:HyperparametersVsIteration}, we find that a good proxy for an event's contribution is $\rho^2 /
m_{2,\mathrm{source}}$.  In a single-event analysis, the measurement uncertainty would depend only on the largest
contributing event, whereas a stacking analysis should scale according to the sum of all events. 
For the plotted analytic
scalings, the RMS errors are {\RMSforN} for $n^{-1/2}$, {\RMSforSTATMAX} for $\max[(\rho^2 /
  m_{2,\mathrm{source}})^{-1/2}]$, and {\RMSforSTATTOT} for $\sum (\rho^2 / m_{2,\mathrm{source}})^{-1/2}$.

Similarly, our stacking strategy can be compared to approaches which investigate the maximum NS mass independently of
the low-mass equation of state.  For a uniformly distributed population with unknown upper limit, ignoring measurement
uncertainty, the upper limit can
be estimated with a statistical uncertainty $\Delta M/N_{big}$ (e.g., via the largest single element), where
$N_{big}$ characterizes the observed number of massive sources and $\Delta M$ characterizes the mass range. 
The top panel
of {Fig.}~\ref{fig:results:HyperparametersVsIteration} illustrates how
our results compare with  such an approach.  
Based on our detection-weighted population parameters, we adopt the scaling $N_{big}=N$, corresponding
to the detection-weighted fraction of
sources associated with the more massive poulation.

\subsection{Understanding Mass distribution constraints and population-reweighted posteriors}

GW measurements will very rapidly identify the chirp mass distribution of merging NS.
As an example, if all NS in merging BNS are drawn from a Gaussian mass distribution, 
then the mean $\bar{m}$ and width $\sigma_m$ of that Gaussian will be identified with confident chirp mass measurements alone to within roughly
 $2.26\sigma/\sqrt{n}$ and
$2.5\sigma/\sqrt{n}$ respectively, using classical frequentist statistics.
This rapid convergence occurs because BNS chirp mass measurements for coalescing binaries with EM counterparts have
statistical errors far smaller than $\sigma$.  The added statistical uncertainty in the absence of NS counterparts only
modestly increases the number of measurements needed for reliable assessment.
Of course, the BNS mass distribution need not be purely Gaussian.  However, if the mass distribution is (for example) a
mixture of distinguishable Gaussians, then similar arguments apply to each component.

The above analysis likewise need not assume all NS are drawn independently from the same distribution.  Indeed,  the paired masses of binary neutron stars could be strongly correlated  through the astrophysical
channels which form them.  But barring astrophysical coincidences,  generic distributions $p(m_1,m_2)$ will  also be constrained by
high-precision one-dimensional chirp mass measurements,  assuming $p(m_1,m_2)$ must be smooth in
$m_1,m_2$ (and not $\mc,q$). %
Above and beyond chirp mass constraints, GW observations also provide direct insight into each individual $q$, albeit
weakly.  For context, for our fiducial unimodal Gaussian mass distribution, the inferred mass ratio distribution is approximately a one-side
normal distribution with mean $q=1$ and width $0.1$---a scale roughly 2/3 of the measurement errors on $q$ expected from
typical PE on our events. 
Therefore qualitatively speaking and pessimistically assuming we must rely
only on mass ratio measurements and not chirp mass, the mean and variance of a presumed
Gaussian mass ratio distribution will converge as roughly $2.26\times (0.1/\sqrt{n})$---modestly more slowly than chirp
mass measurements alone will constrain the width.

\begin{figure*}[htbp]
\centering
\includegraphics[width=\textwidth]{%
  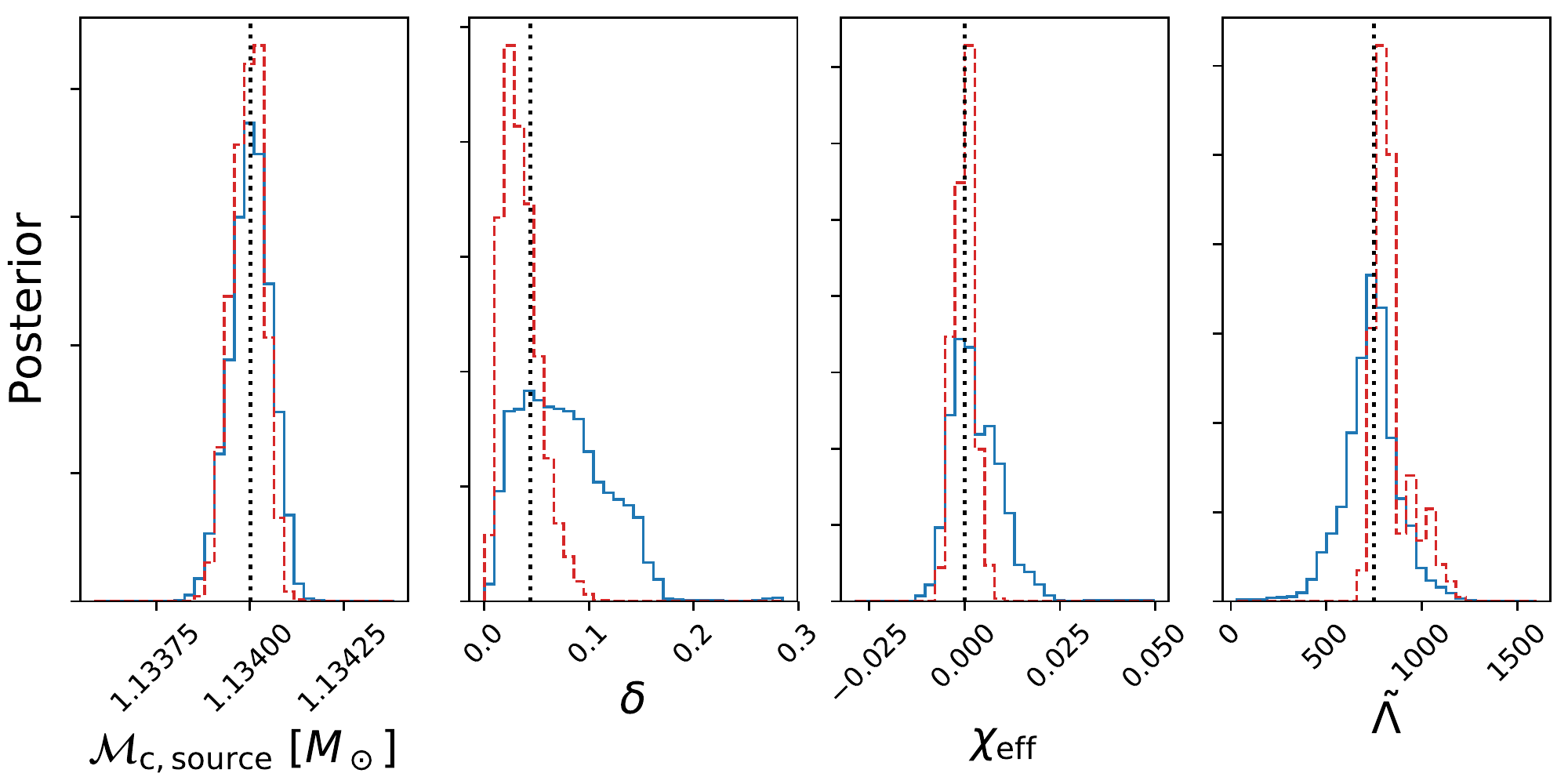%
}
\caption{
Population-informed priors:
posterior distribution for a moderately loud synthetic BNS using a prior uniform
in the plotted variables (solid blue) and a population-informed prior (dashed
red), wherein the events used to constrain the population do not include the
event plotted.  Notice the population-informed results are much more precise,
and zoom in on the injected parameters (shown as vertical black lines).
}
\label{fig:reweighted-event}
\end{figure*}

One way or another---via chirp mass constraints or direct constraints on the mass ratio distribution from stacked individual events---our inferences
about the population's mass ratio distribution should significantly decrease the expected uncertainty in $q$ for future
observations.  As a concrete example, Figure \ref{fig:reweighted-event} shows the result of interpreting a significant-amplitude 11th
event after first observing 10 NS mergers from our synthetic population.  The inferred mass ratio constraints are substantially tighter.
We also show the joint posterior distribution in masses, spin, and tides for this new event.   Using
population-informed priors for the mass ratio distribution, we draw tighter conclusions about the new  events' potential
tidal deformability. %

\begin{figure}
\includegraphics[width=\columnwidth]{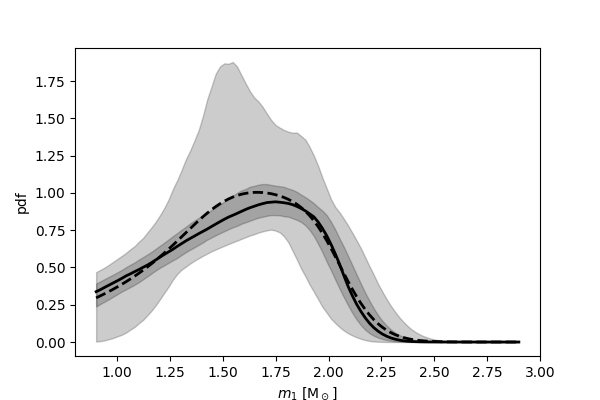}
\caption{
  Impact of unsuitable mass distribution model:
  Median (solid line), posterior predictive distribution (dashed line), 50\% and
  90\% credible regions (shaded) for $m_1$, adopting a unimodal model for
  inference on our synthetic data.  Compare to the right panel of
  {Fig.}~\ref{fig:results}.
}
\label{fig:ppd:m1_unimodal}
\end{figure}

Our choice of  NS mass distribution model can significantly impact our inferences.  As an example,  Figure \ref{fig:results:eos_central} shows
the results of inference using a unimodal NS mass distribution.   As shown in Figure \ref{fig:ppd:m1_unimodal}, at small $n$ this poorly-fitting model would suggest
the maximum mass is significantly constrained by the absence of high-mass observations, as a single very wide Gaussian
would be required to match the mean and dispersion of our two-component model.  
Our choice of mass model has a substantial impact on the inferred equation of state.
We emphasize that the mass and spin distribution  is observationally constrained by the many low-amplitude observations
for which tides are largely inaccessible; therefore, it's important to use  all observations to produce an
unbiased estimate of the EOS.

Do we need to simultaneously constrain the nuclear EOS and the NS mass distribution, ignoring spin?  For our scenario, chirp mass measurements
alone dominate our ability to recover the mass distribution.  We therefore do not expect joint inference to
significantly alter the small-$n$ results: we could alternatively  first estimate the NS mass distribution,
and then reconstruct the inferred nuclear EOSwith care to not double-count the candidate event's likelihood.

\subsection{Recovering the spin distribution}

\begin{figure}[htbp]
\includegraphics[width=\columnwidth]{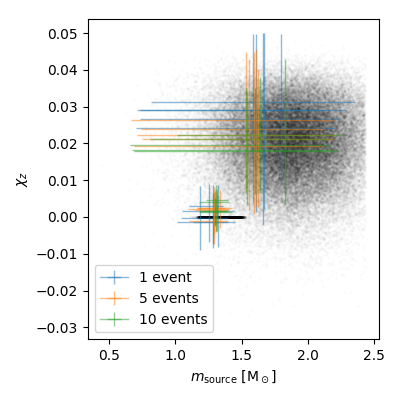}
\caption{
Posterior predictive distributions for the two-component mixtures estimated with
$1$, $5$, and $10$ BNS observations.  The markers indicate the recovered mean and standard deviation for each component.
}
\label{fig:ppd:mass_spin}
\end{figure}

Our synthetic population has zero NS spin for one component, and observationally accessible NS spin for the more massive
component.    As illustrated by Figure \ref{fig:ppd:mass_spin}, we can therefore recover the
joint mass and spin distribution of each component with very few observations.  
As with the mass distribution, we have intentionally adopted a model -- both NS in a binary drawn from the same mass and
spin distribution -- which is more easily constrained by GW observations, to highlight the impact of joint mass-spin
distribution constraints  on the EOS.

\begin{figure*}[htbp]
\centering
\includegraphics[width=\textwidth]{%
  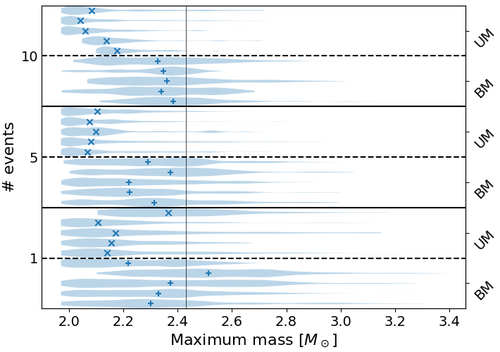%
}
\caption{
Mass distribution impact on EOS upper mass limit:
Posterior distributions for the upper mass limit, estimated with 5 random
realizations each of 1, 5, and 10 BNS observations.  True values for injected
EOS shown as vertical bars.  Separate re-runs were done with both the accurate
bimodal (BM) mass distribution and the biased unimodal (UM) mass distribution.
Bimodal model slowly converges to tighter constraints around the correct value,
whereas the unimodal model rapidly converges to a biased value, emphasizing the
need for careful mass distribution modeling.
}
\label{fig:post:mass_ranges}
\end{figure*}

\SkipPreliminary{
Do we need to simultaneously constrain the nuclear EOS and the NS mass and spin distribution?  
Tides and NS spin have a correlated impact on the radiated signal.   We therefore expect combined
inference of mass, spin, and tides will be required.  To illustrate the need for NS tides, Figure  \ref{fig:ppd:spin_test_notides} shows the inferred 
spin distribution,   assuming exactly zero tides for all NS.

\begin{figure*}[htbp]
\centering
\includegraphics[width=\textwidth]{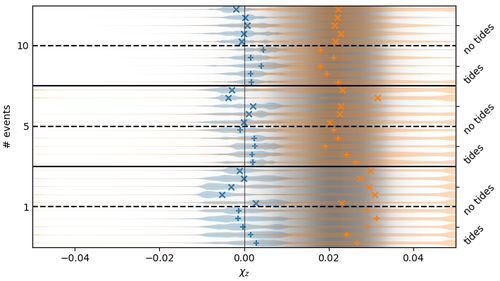}
\caption{\label{fig:ppd:spin_test_notides}
Tidal impact on spin recovery:
Posterior predictive distributions for spin magnitudes under the two-component
mixture model, estimated with 5 random realizations each of 1, 5, and 10 BNS
observations.  Separate re-runs done where parameter estimation incorrectly
assumes vacuum solutions (i.e., $\Lambda_1 = \Lambda_2 = 0$), labeled ``no
tides,'' to emphasize improved constraints when spin-tide correlations accounted
for.  True population overlayed as grey densities.

}
\label{fig:ppd:spin-tide-effects}
\end{figure*}
}

\section{Discussion}
\label{sec:discussion}

In this work, using a concrete but extreme synthetic example, we demonstrated that the whole merging NS population provides vital insight into constraining  the  NS EOS.
The faint events tell us the NS mass (and spin) distribution.  Using that information, we better interpret the loud
events' masses, drawing sharper conclusions on the NS EOS. 
We furthermore demonstrated that the EOS must be simultaneously inferred along with the NS mass and spin distribution,
to avoid introducing bias.  
In this section, we highlight areas in which our synthetic example might not be representative, while presenting how the
lessons learned from it should translate to more realistic future observing scenarios.

First and foremost, we emphasize we have made one key extreme assumption to allow us to highlight the contribution from constraints on the NS
maximum mass: we assume the second component is comprised of massive NS's which are rapidly spinning.  In some
formation scenarios for high-mass NS, the massive NS accretes substantial matter (and spin) through CE accretion \cite{2005ApJ...632.1035O,2013A&ARv..21...59I,2014LRR....17....3P}.  We would therefore more likely expect massive, rapidly-rotating NS to be paired with low-mass
companions.  Instead, our straw man model  produces binaries with well-measured chirp masses near the maximum value
allowed by our EOS, enabling sharper constraints than would be expected from scenarios with mixed NS binaries.

Our EOS models lack phase transitions and thus imply  strong correlations between the maximum mass $m_{\rm max}$ and
tidal deformability.%
These two measurements therefore provide two avenues to constrain the nuclear
  equation of state.   If we adopted a more flexible model for the nuclear equation of state, our implicit use of two observables
  (maximum mass and tides) would not necessarily enable relatively tighter  constraints on the EOS than the use of each observable independently.

Similarly but on a longer timescale, GW measurements will gradually pin down the BNS spin distribution, as observations
accumulate enough in number to probe at and significantly below the measurement error of the loudest expected signals.
  After the first few measurements, the mass ratio distribution could be  very
strongly constrained,  confidently disfavor highly asymmetric binaries, and therefore substantially decrease statistical
uncertainty in spin.   At that level, the statistical uncertainty in spin will be of order $0.01$, which could be
produced by astrophysical formation channels. 
  A population of NS spins consistent with zero is  plausible, easily
tested, and simplifies the discussion we continue below. However, because of the correlation of spin and tides, if spins are
nonzero, the distribution in EOS and spin must be carried out together.

This information from the low-significance population helps inform the interpretation of the roughly one in ten BNS
mergers with amplitude $\rho>20$ which provide the most information individually about the EOS.    The mass ratio, spin,
and  chirp mass are all correlated with the inferred tidal deformability $\tilde{\Lambda}$.
Because we can better constrain each individual measurement, we draw more information about tides with each observation
when we use joint inference.  The degree of advantage depends on the astrophysical NS mass and spin distribution and the
EOS; as we've shown, a distribution extending close to the maximum mass can be very informative.

\section{Conclusions}
\label{sec:conclusions}

As demonstrated by the direct detection of gravitational waves (GWs) from neutron stars and black holes  \cite{LIGO-GW170817-SourceProperties,LIGO-O2-Catalog}, the universe naturally provides a highly-relativistic ``cosmic collider'' for pairs of compact objects---black holes
(BHs) or neutron stars (NS).  For each collision, current and future GW observations can identify the nature of the
coalescing binaries, the dynamics of the collision, and the nature of the post-merger remnant \cite{LIGO-Program-2019}, providing direct insight into the physics of each merger.   Moreover, the population of observations will
enable direct measurements of the population of merging binaries themselves---their joint mass, spin,  redshift, and eccentricity distribution.
In this work, we demonstrate one use  of this cosmic collider: joint inference about the phenomenological astrophysical distribution of
merging NS properties (mass and spin) simultaneously with the nuclear equation of state. 
Analyzing a fixed ensemble of synthetic data,   we show that joint inference of the NS mass, spin, and tides with all
NS observations are
required to reliably infer the EOS.  We in particular demonstrate that even low-amplitude NS observations contribute
significantly to constraining the NS, albeit indirectly, by providing strong constraints on the NS mass and spin
distribution.  By contrast, previous work has argued that all information about the NS EOS is carried in the most
massive observations.   We demonstrate significant biases could occur if the mass distribution is
inappropriately modeled.  
And we reviewed how NS observations will rapidly constrain the NS mass and spin distribution.     Our concordance approach can be immediately generalized to incorporate other observational
  constraints, extending other similar work which assumes the NS mass and spin distribution is known (e.g., \cite{2019arXiv190408907M}).
The   \textsc{PopModels} code is publicly available \cite{gwastro-PopulationReconstruct-Code-PopModels}, as are all
information used to reproduce the examples in this work.

In this proof-of-concept work, we adopted several strong assumptions about the NS population, to enhance the impact that
joint inference has on the inferred EOS.   Notably, we  assumed the NS mass distribution extended to the maximum mass
supported by the equation of state.   Also, motivated by galactic observations, we also did not introduce an extended
population of asymmetric NS binaries.   A more comprehensive analysis of real observations should relax both
assumptions.

In this paper, we only illustrated a few scenarios for future GW observations, assuming a relatively simple population
of unambiguous binary neutron stars.   While we do not address a closely related question---distinguishing between
populations of BH-NS and NS-NS (and BH-BH) of similar mass---our concordance framework provides a natural framework
within which to address this question.  It will immediately allow for  multiple populations, incorporate populations with exactly
zero tides, and directly employ the correct likelihood normalization (i.e., evidence) into all 
calculations.  
We will more carefully investigate the question of multiple compact binary populations with similar mass in future work.

As observations accumulate, our ability to identify the nuclear EOS can   be increasingly impacted by systematic biases
in our understanding of GR \cite{2014PhRvL.112j1101F,2014PhRvD..89j3012W,2015PhRvD..91d3002L,2018PhRvD..98l4030S,2019arXiv190912718K}, barring steadily-increasing model accuracy as
in
\cite{2015PhRvL.115i1101B,2016PhRvL.116r1101H,2018PhRvD..97d4044K,2017PhRvD..96l1501D,2019PhRvD..99b4029D,2018PhRvD..98j4052N,2019arXiv190409558M}.
Our inferences about the EOS  can also be influenced by biases or inflexibility in our EOS parameterization;  see,
e.g.,  \cite{2019MNRAS.485.5363G}.
 Using RIFT or other efficient parameter inference engines to draw conclusions about individual
events using different waveform models with different systematics, one can directly assess
the impact of these systematic errors on the inferred population and EOS.

Though electromagnetic observations of galactic pulsars and binary mergers provide a complementary avenue to constrain the
nuclear equation of state, the tightest constraints in the future will exploit all messengers.  
Some investigations have already jointly constrained the EOS by combining galactic  X-ray binary observations with
GW170817 tidal constraints \cite{2019arXiv190205078F,2019arXiv190204557K}. %
Another promising approach attempted with GW170817 proposes to identify the nature of the postmerger object from the
presence (or absence) of electromagnetic emission \cite{2017ApJ...850L..19M,2017ApJ...850L..34B,2018ApJ...852L..29R}.  Large-scale statistics on even qualitative features of  remnants can
inform the EOS \cite{2019-MargalitMetzger-ChirpMassAndEMconstraintsEOS}, though the efficiency and utility of such
qualitative stacking depends strongly on followup EM surveys, on systematic biases or substantial theoretical uncertainties associated with the interpretation of individual
EM measurements (e.g. \cite{2019arXiv190301466K}), and on theoretical modeling uncertainty associated with the transitions between the different proposed
postmerger scenarios.
Conversely, as the amount and nature of the ejected material depends strongly and delicately on the merger's binary
parameters (e.g., mass ratio and spin), the same population-modeling techniques described in this work will also need to
be applied to interpret electromagnetic observations too  (e.g. \cite{2018ApJ...866...60P,2019arXiv190301466K}).
We defer  discussion of any multimessenger constraints to future work.

While we defer exporations of other applications to future work, the method described here will quickly translate to
other applications which exploit the simultaneous interpretation of multiple coalescing NS.
For example, gravitational wave measurements of coalescing compact binaries can also be used as standard candles, to help inform the
cosmic distance ladder  \cite{2010ApJ...725..496N,2013arXiv1307.2638N,2018arXiv181111723M,2017Natur.551...85A,2018Natur.562..545C}.     As  GW measurements alone constrain the luminosity distance $d_L$ and
redshifted mases $m_{i,z}=m_i(1+z)$, cosmological constraints require a third constraint: some independent constraint
(e.g., a host galaxy or preferred length scale), providing access to either $m_i$ or $z$ and therefore enabling
cosmological constraints.   Even without observational counterparts, binary neutron stars may have distinctive mass
\cite{2012PhRvD..85b3535T} and tidal features whose observation could potentially enable better cosmological constraints (e.g.,
\cite{2012PhRvL.108i1101M,2017PhRvD..95d3502D}).

The strategy in this work relies on accurate likelihood estimates $\ell_n(x)$, provided through RIFT and libraries used
therein.
Conventional machine learning
 techniques can provide very accurate universal  function approximations with feedforward neural
networks; see, e.g., \cite{Cybenko-UniversalApproximationTheorem,Hanin-UniversalApproximationsPractical}. 
Discussion of alternative interpolation techniques will be presented in a forthcoming publication. %

Given the substantial astrophysical and modeling uncertainties involved, we have employed a  phenomenological approach.
We recognize that strong prior assumptions about the NS population or equation of state could provide stronger and more
rapid (conditional) constraints, and we defer to substantial prior work in this area for a discussion of the relevant
techniques and pitfalls
 \cite{2015ApJ...810...58S,2018PhRvD..97d3014W,2018PhRvD..98h3017T}.

\begin{acknowledgments}
The authors thank Will Farr for helpful suggestions for future work.  
ROS and DW gratefully acknowledge NSF award PHY-1707965 and AST-1909534.
DW also acknowledges support from RIT through the FGWA SIRA initiative.
Computational resources provided by the LIGO Laboratory and supported by National
Science Foundation Grants PHY-0757058 and PHY-0823459 are gratefully acknowledged.
\end{acknowledgments}

\appendix

\section{Scaling accuracy with increasing measurements: a Fisher perspective}
\label{ap:scaling}
In the text, we provide a concrete illustration of how well we can measure the nuclear equation of state given several
coalescing binary neutron star measurements, using all available information and employing phenomenological models for
the NS population and the EOS.  We find that the added information from
low-significance events better constrains the mass and spin distribution; when applying this insight to the loudest
signals, these low-significance events thereby help indirectly further constrain the nuclear equation of state.

In this appendix, to facilitate projections to future instruments and other observational scenarions, we provide a more
qualitative outline of this argument using Fisher matrix methods.  While we frame our discusion using the terminology of
nuclear equation of state measurements, our discussion is not specific to that case.

In the local universe, the amplitude distribution for confidently-identified sources will be nearly Euclidean, with the fraction of sources with
network amplitudes greater than $\rho$ determined by $P(>\rho)=(\rho_{min}/\rho)^3$, where $\rho_{min}$ is some minimum
identifiable amplitude.   Only a subset of parameters will be accessible for signals near the detection threshold.   For
sufficiently loud signals $\rho>\rho_{cut}$, however, additional features of the coalescing binary will be apparent---for the purposes of this discusison, the effective binary tidal deformability $\tilde{\Lambda}$.  In this discussion we
will adopt $\rho_{min}=10$ and $\rho_{cut}=20$.  With these assumptions, out of $N$ sources, on average only  $N/8$ will provide
information about tides and therefore provide enough information in isolation to produce any constraint on the nuclear
equation of state. 
Another important quantity is $\E{\rho^2} = \int d\rho \rho^2 dP/d\rho  = 3\rho_{\rm min}^2$, so for a sum over $N$
sources, the average value of $\sum_k \rho_k^2 $ is approximately $3N \rho_{\rm min}^2$.

The non-marginalized likelihood in the full $Nd+1+D $-dimensional space of all binary parameters and all population
hyperparameters is the integrand appearing in Eq. (\ref{eq:inhomog-poisson-likelihood}): $\ln L = - \mu + N \ln {\cal R}
+ \sum_n \ln \ell_n(x_n) + \ln p(x_n|X)$, where $x_n$ are $d$-dimensional variables characterizing each event.  
More broadly, the  likelihood $\ln L$ can be expanded in a Taylor series in $y=(x_1\ldots x_n,{\cal R}, X)$
around its maximum:
\begin{align}
\ln L = \text{const} - \frac{1}{2} \Gamma_{\alpha \beta} (y-y_*)_\alpha (y-y_*)_\beta
\end{align}
Constraints on the EOS follow by marginalizing this likelihood over all parameters except the subset of $X$ that
corresponds to the EOS.   When carrying out this calculation, we want to qualitatively assess how much we learn about
the EOS by exploiting better constraints on the mass distribution, particularly as provided by the weak sources which
don't independently inform the EOS.

To provide qualitative insight into this marginalization, we first break up the components in Taylor series themselves.
We assume that in
suitable coordinates, the individual likelihoods $\ell_n$ are nearly Gaussian for variables which are well constrained,
and nearly constant for poorly-constrained variables; in the context of this discussion, the variables $\mc, \eta,
\chi_{\rm eff}$ are assumed to be well-constrained always, with $\tilde{\Lambda}$ constrained  for strong sources: that
is,
\begin{align}
\ln \ell_n(x) \simeq - \frac{1}{2} \rho_n^2 \hat{\gamma}_{ab}^{(k)} (x-x_{n,*})_a(x-x_{n,*})_b + \text{constant}
\end{align}
where $\rho_n$ is the amplitude of the $n$'th source.
Moreover, to simplify our argument, we will assume $\hat{\gamma}_{ab}$ is \emph{independent} of binary parameters, and exists
in one of two classes: the ``strong'' sources (S) which constrain the added tidal parameters of interest, and the ``weak''
sources (W), for which these parameters remain unconstrained.

We first consider a simplified scenario where  the model hyperparameter $X$ we seek to constrain is in fact one of our observables
$x$ for the individual NS observations---in our scenario, for example, all NS could have a   common
radius $R_{ns}$ and are drawn from a common Gaussian mass distribution with unknown mean $\bar{m}$, but our ability to measure that radius could be correlated with other binary parameters like the NS mass.   After
integrating out the deterministic relationship between $x_n$ and $X$, and omitting the event rate ${\cal R}$ and
sensitivity $\mu$ as superfluous, we end up with an expression 
\[
\ln L \simeq \text{const} - \frac{1}{2} \sum_k \rho_k^2\hat{\gamma}_{ab}(x - x_{k,*})_a(x - x_{k,*})_b
\]
where $x=(\bar{m},R_{ns})$ now characterizes the parameters held in common and $x_{*,k}$ characterize the specific
choices which maximize the likelihood for each individual event.   The signal amplitude $\rho_k$ and signal parameters $x_{*,k}$ are uncorrelated.  Therefore, in this expression, we
naturally find two types of terms appearing naturally:
\begin{align}
\sum_{k\in S} \rho^2 \hat{\gamma}^{S}_{ab} &= \hat{\gamma}^{S}_{ab}  3 N \rho_{\rm min}^2 P(>\rho_{cut}) 
 =  \hat{\gamma}^{S}_{ab} 3 N \frac{\rho_{\rm    min}^5}{\rho_{cut}^3} \\
\sum_{k\in W} \rho^2 \hat{\gamma}^{W}_{ab} &= 3 N \rho_{\rm min}^2 \hat{\gamma}^W [ 1- P(>\rho_{cut})]
\end{align}
and thus the likelihood can be approximated up to an overall constant as
\begin{align}
-\frac{2}{3\rho_{\rm min}^2} \ln L &\simeq  \hat{\gamma}^{(W)}_{ab}\sum_{k\in W}(x -x_{k,*})_a(x-x_{k,*})_b  \\
 & +\hat{\gamma}_{ab}^{(S)} P(>\rho_{cut}) \sum_{k\in S}(x-x_{k,*})_a(x-x_{k,*})_b  \nonumber
\end{align}
Within the context of this subsection, $\hat{\gamma}^{W}_{ab}$ has only one nonzero term, for the mass component, while
$\hat{\gamma}^{S}_{ab}$ has all three components nonzero.
The second term reflects how a few strong signals provide information about the hard-to-measure parameters like $R_{ns}$.
The first term reflects how many weak measurements provide information about the NS mass distribution in general and the
mean NS mass $\bar{n}$ in particular, but not
hard-to-constrain parameters like $R_{ns}$.    However, by providing additional information about the NS mass, they \emph{can} help support
constrain the remaining hyperparameters.  In this concrete scenario, the parameters $\bar{m},R_{ns}$ have a statistical
covariance (squared measurement error) of
\begin{eqnarray}
\Sigma = \frac{1}{3\rho_{\rm min}^2 N} [ \hat{\gamma}^W +  P(>\rho_{cut}) \hat{\gamma}^S]^{-1} 
\end{eqnarray}
If significant correlations exist between $R$ and $\bar{m}$, then the measurement accuracy for $\bar{R}$ when we
simultaneously constrain $R,\bar{m}$ can be
noticably  smaller.   Additional correlations provide additional opportunities for improvement.

\section{Improved EOS coordinate system with PCA}
\label{ap:pca}

The pressure-based spectral parameterization for neutron star equation of state has an issue in that its parameters
$\gamma_0$, \ldots, $\gamma_n$ are only physical in a small subspace, which is not aligned with the coordinate axes.  
Since we want to reject any point with large $\max_{p} \Gamma$, our priors are not well-suited to our  choice of
basis functions.  For example, using  a Legendre polynomial basis  $\Gamma(x)=\sum_k P_k(x) \bar{\gamma}_k$ insures a
bound on $\gamma$ is related to a bound on $\Gamma$.  
Still, any method which draws random $\gamma_i$ samples is going to have to deal with the tight correlations.  To deal with this issue, we consider the general problem of an $n$ dimensional volume $\mathscr{V}$ enclosed in a hypercube $\mathscr{C}$, where $\mathscr{C}$ is known analytically, but $\mathscr{V}$ is only known by a procedure which can determine if a point $\mathscr{P}$ is contained in $\mathscr{V}$.  Our goal is to find the minimal hypercube $\mathscr{C}'$ which encloses $\mathscr{V}$.  In our specific EOS example, $\mathscr{C}$ is the 4 dimensional hypercube of spectral EOS parameters---bounded by $\gamma_0 \in [+0.20, +2.00]$, $\gamma_1 \in [-1.60, +1.70]$, $\gamma_2 \in [-0.60, +0.60]$, $\gamma_3 \in [-0.02, +0.02]$.  $\mathcal{V}$ is the subset of $\mathscr{C}$ which define equations of state permitted by physics.  From a Monte Carlo study, we find that $\mathcal{V}$ comprises $\simeq \plaineff$ the volume of $\mathscr{C}$, and thus any procedure which draws random samples uniformly in $\mathscr{C}$ will have one in 20000 be physical.

To find $\mathscr{C}'$, we start by drawing samples from $\mathscr{C}$ until we have found $N$ within $\mathscr{V}$ (here $N = \pcansamples$).  Let's call a sample in the basis aligned with $\mathscr{C}$ ``$\boldsymbol{r}$.''  Now we rescale all of these samples by subtracting the sample mean vector $\boldsymbol{\mu}_r$, and dividing component-wise by the sample standard deviation vector $\boldsymbol{\sigma}_r$
\begin{equation}
  \tilde{\boldsymbol{r}} =
  (\boldsymbol{r} - \boldsymbol{\mu}_r) / \boldsymbol{\sigma}_r.
\end{equation}
We can then feed these standardized $\tilde{\boldsymbol{r}}$ samples into a principal component analysis (PCA) routine (in this case provided by \textsc{scikit-learn}'s \texttt{sklearn.decomposition.PCA} class \cite{scikit-learn}).  This method finds a rotated coordinate system, $\boldsymbol{r}'$, in which the first dimension captures the majority of the data's variance, and each subsequent orthogonal dimension captures the majority of the remaining variance.  The transformation from $\tilde{\boldsymbol{r}}$ to $\boldsymbol{r}'$ is encompassed in a matrix operator $\boldsymbol{S}$, in which each row contains the components of the $\boldsymbol{r}'$ bases in the $\tilde{\boldsymbol{r}}$ coordinate system, such that
\begin{equation}
  \boldsymbol{r}' = \boldsymbol{S} \, \tilde{\boldsymbol{r}}.
\end{equation}
In this $\boldsymbol{r}'$ coordinate system, we compute the minimum and maximum values of each sample in each dimension, which combined make the boundaries of our more efficient hypercube, $\mathscr{C}'$.  In the case of our EOS parameters, sampling uniformly within $\mathscr{C}'$ provides us with an efficiency of $\simeq\pcazeroeff$, $\pcazerologimp$ orders of magnitude better.  However, due to the limited sample size used to find $\mathscr{C}'$, it is possible that a small portion of $\mathcal{V}$ is outside of $\mathscr{C}'$.  To reduce the odds of this, $\mathscr{C}'$ can be enlarged to include some buffer space.  We employ a simple strategy here, by extending the hypercube by an additional $10\%$ in each direction.  This can be adjusted according to one's tolerance needs.  In this extended $\mathscr{C}'$, our efficiency is $\simeq\pcateneff$, which corresponds still to a $\pcatenlogimp$ order of magnitude improvement.

\begin{figure}[htbp]
  \centering
  \includegraphics[width=\columnwidth]{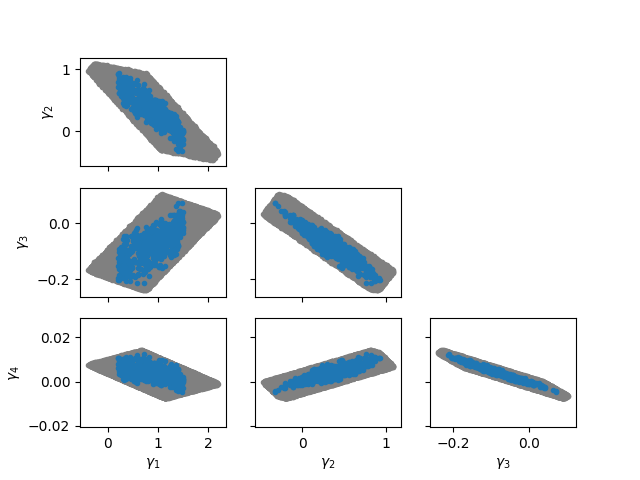}
  \caption{%
    Valid samples used in fitting (blue) and projections of the
    bounding hypercube (gray), for each pair of spectral coordinates.  We
    show the bounds only for the non-expanded $\mathscr{C}'$ here.
  }
  \label{fig:samples-and-hypercube}
\end{figure}

See Fig.~\ref{fig:samples-and-hypercube} for the fit used, Table \ref{tab:transformation} for the components of the transformations, and Table \ref{tab:bounds} for the non-buffered hypercube bounds in the transformed coordinates. %

\begin{table}[htbp]
  \centering
  \include{tables/pca-table}
  \caption{Transformation components for $\boldsymbol{r} \to \boldsymbol{r}'$.}
  \label{tab:transformation}
\end{table}

\begin{table}[htbp]
  \centering
  \include{tables/hypercube-bounds}
  \caption{Hypercube bounds for the non-expanded $\mathscr{C}'$.}
  \label{tab:bounds}
\end{table}

\bibliography{paperexport,LIGO-publications,eos-constraints-papers}

\end{document}

%% file: tables/pca-table.tex
\begin{tabular}{llll|l|l}
\multicolumn{4}{c}{$\boldsymbol{S}$} & \multicolumn{1}{c}{$\boldsymbol{\mu}_r$} & \multicolumn{1}{c}{$\boldsymbol{\sigma}_r$} \\ \hline
$+0.43801$ & $-0.53573$ & $+0.52661$ & $-0.49379$ & $+0.89421$ & $+0.35700$ \\
$-0.76705$ & $+0.17169$ & $+0.31255$ & $-0.53336$ & $+0.33878$ & $+0.25769$ \\
$+0.45143$ & $+0.67967$ & $-0.19454$ & $-0.54443$ & $-0.07894$ & $+0.05452$ \\
$+0.12646$ & $+0.47070$ & $+0.76626$ & $+0.41868$ & $+0.00393$ & $+0.00312$ \\
\end{tabular}

%% file: tables/hypercube-bounds.tex
\begin{tabular}{l|llll}
& \multicolumn{1}{c}{$r'_0$} & \multicolumn{1}{c}{$r'_1$} & \multicolumn{1}{c}{$r'_2$} & \multicolumn{1}{c}{$r'_3$} \\ \hline
min & $-4.37722$ & $-1.82240$ & $-0.32445$ & $-0.09529$ \\
max & $+4.91227$ & $+2.06387$ & $+0.36469$ & $+0.11046$ \\
\end{tabular}